# Improved fragment-based movement with LRFragLib for all-atom Ab initio protein folding

Tong Wang[1,2], Haipeng Gong[1,2,*] and Eugene I. Shakhnovich[3,*]

[1]MOE key laboratory of bioinformatics, School of Life Sciences, Tsinghua University, Beijing 100084, China, [2]Beijing Innovation Center of Structural Biology, Tsinghua University, Beijing 100084, China, [3]Department of Chemistry and Chemical Biology, Harvard University, Cambridge, Massachusetts, United States of America

*To whom correspondence should be addressed.

## Abstract

Fragment-based assembly has been widely used in Ab initio protein folding simulation which can effectively reduce the conformational space and thus accelerate sampling. The efficiency of fragment-based movement as well as the quality of fragment library determine whether the folding process can lead the free energy landscape to the global minimum and help the protein to reach near-native folded state. We designed an improved fragment-based movement, "fragmove", which substituted multiple backbone dihedral angles in every simulation step. This movement strategy was derived from the fragment library generated by LRFragLib, an effective fragment detection algorithm using logistic regression model. We show in replica exchange Monte Carlo (REMC) simulation that "fragmove", when compared with a set of existing movements in REMC, shows significant improved ability at increasing secondary and tertiary predicted model accuracy by 11.24% and 17.98%, respectively and reaching energy minima decreased by 5.72%. Our results demonstrates that this improved movement is more powerful to guide proteins faster to low energy regions of conformational space and promote folding efficiency and predicted model accuracy.

# 1 Introduction

Ab initio protein folding simulations have been widely used to detect folding pathways and to predict protein three dimensional structures (Bradley, et al., 2005; Dill and MacCallum, 2012; Lee, et al., 2017; Soding, 2017; Yang, et al., 2007). During the folding process, the free energy has a rugged funnel-like landscape biased toward the native structure (Onuchic, et al., 1997; Wolynes, 2015), where, in principle, the native state is characterized by the free energy minimum state that can be detected by minimization of energy functions(Schafer, et al., 2014). Although energy functions has been well developed in decades, the accuracy of protein structure prediction was still hindered by the enormous of degrees of freedom of protein conformations (Jothi, 2012; Kim, et al., 2009). Fragment-based assembly is an accessible approach to reduce the conformational space and accelerate sampling. Exemplified by 3-mer and 9-mer fragment-based move set in Rosetta, this approach has been widely used in most ab initio protein folding simulations (Simons, et al., 1997; Xu and Zhang, 2012; Yang, et al., 2015), although non-fragment based move sets are also used together and play an important role in the folding process (Chowdhury, et al., 2003; Duan and Kollman, 1998; Ołdziej, et al., 2005; Ozkan, et al., 2007; Simmerling, et al., 2002; Srinivasan and Rose, 2002; Tian, et al., 2016). During the simulation, fragment-based movements sample the conformational space by repetitively substituting torsion angles of the target protein with identified short fragment structures derived from a fragment library constructed by other algorithms (Bonneau and Baker, 2001). Since fragment-based movements consider the genomic constraints resulting from the arrangements of amino acids in the primary sequences, it greatly limits conformational searching space and thus improves the folding efficiency (Wang, et al., 2017).

The quality of fragment library is fundamental to the efficiency of fragment-based movement. State-of-the-art fragment library construction algorithms such as NNMake (Gront, et al., 2011), HHfrag (Kalev and Habeck, 2011), SAFrag (Shen, et al., 2013) and

Flib (de Oliveira, et al., 2015) have been proposed and utilized into protein structure prediction program. More recently, our group designed an effective algorithm, LRFragLib (Wang, et al., 2017), which outperforms existing approaches by achieving a significantly higher precision and a comparable coverage in sampling near-native structures. Based on two kinds of logistic regression models, LRFragLib utilizes a mutil-stage, flexible selection protocol to detect near-native fragments of 7-10 residues. In this study, we consider the discretization of the natural coordinates of the peptide backbone, which are the $\Phi$ and $\Psi$ angles. First, we used LRFragLib to generate a fragment library containing $\Phi$ and $\Psi$ angles derived from known fragments for the target protein. Then we randomly picked out one from the fragment library as the template and substituted all torsion angles of a chosen position of the target protein during each step of replica exchange Monte Carlo (REMC) simulation (Yang, et al., 2007).

REMC simulation is an all-atom ab initio protein folding program designed by our previous research. With a set of movements and well developed energy functions, REMC can lead small proteins within 100 residues to fold into their near-native structures. Three backbone movements and a side chain movement are adopted into REMC, in which the global and the local move both make torsion angle substitutions based on the randomly selected values, while the knowledge-based move which only changes one pair of $\Phi$ and $\Psi$ angles in one step. Different from all these backbone movements, we designed a movement to substitute 7-10 pairs (corresponding to the length of chosen template fragments) of $\Phi$ and $\Psi$ angles of the target protein with the corresponding angles of the template with an optimized noise range in each simulation step. We call this movement the "fragmove", since the choice of pairs of torsion angles from a fragment library of known structures. Considering that "fragmove" provides a finite collection of dihedral pair values and using this move set alone in a folding simulation would be too much restrictive, we combined "fragmove" with other backbone movements in REMC with optimized weights. Finally, we systematically analyzed the performance of REMC simulations with or without fragmove both on the testing set used in our previous research and the 12th Critical Assessment of Structure Prediction (CASP12) set (Kryshtafovych, et al., 2016).

**2 Methods**

2.1 Datasets

The training and testing sets are the same to the datasets utilized in previous REMC research (Yang, et al., 2007). Five proteins with different folds and lengths ranging from 39 to 60 residues were used for the training set (Table S1) while twelve proteins with lengths ranging from 40 to 77 residues were used for the testing set (Table S2). Besides the testing set used before, we also built a CASP12 set to evaluate the performance of REMC with or without fragmove. We extracted all template-free modeling (FM) targets which have full atom information in CASP12 competition. Twelve proteins with the lengths ranging from 55 to 161 residues are included in the CASP12 set (Table S3).

2.2 Fragment library generation and assessment

We used LRFragLib with default parameters to generate fragment library for each target protein (Wang, et al., 2017). For each target protein, all homologous proteins were carefully removed before running LRFragLib. Secondary structure and dihedral angles predicted by SPIDER3 (Heffernan, et al., 2017) were taken as input for LRFragLib program. Here we used two indices, precision and coverage, that were widely accepted to evaluate the quality of fragment library. Precision is defined as the proportion of good fragments while coverage is defined as the proportion of positions that are covered by at least one good fragment, where good fragments are those which are structurally close to the native fragment within an RMSD cutoff value. Same to the evaluation procedure in LRFragLib research, we used a series of RMSD10 cutoffs from 0.1 to 2.0Å, where

RMSD10 is the normalized RMSD value to that of 10-residue fragments (Carugo and Pongor, 2001).

2.3 Fragmove

Fragmove was formulated in the following way: for each position of the target protein that spans 7-10 residue window, we queried all candidate fragments from the fragment library generated by LRFragLib and recorded their 7-10 pairs of $\Phi$ and $\Psi$ angles. During the simulation, fragmove entailed setting the dihedral angles of all residues of a randomly selected position to those of the template. The step size of fragmove was drawn from a Gaussian distribution with the mean of the difference between the torsion angle of the target and that of the template and an optimized standard deviation. Then fragmove was incorporated into replica exchange Monte Carlo simulation with other backbone movements. The weight of using fragmove to make backbone torsion angle substitution as well as the standard deviations of the step sizes of $\Phi$ and $\Psi$ angles were optimized on the training set. Parameters were optimized by minimizing RMSD values and maximizing TM-Score values (Zhang and Skolnick, 2005) of the lowest energy structures.

**Fig. 1.** General flowchart (A) and move sets in each step (B) of replica exchange Monte Carlo (REMC) simulation.

2.4 REMC simulation

An overall flowchart describing the pipeline of REMC is shown in Figure 1. A set of movements and an all-atom knowledge-based statistical energy function are adopted in REMC (Xu, et al., 2011; Yang, et al., 2007; Yang, et al., 2008). The statistical energy function is the sum of pairwise atom-atom contact energy, hydrogen-bonding energy, torsional angle energy and side chain torsional terms, with an additional orientated-dependent term describing nearby aromatic residues. The movements consist of rotations about φ, ψ, and χ dihedral angles of all residues except proline, with bonds and angles held fixed. Including fragmove, there are four kinds of backbone movements and a side chain move adopted in REMC (Figure 1B). A global move is to rotate the Φ or Ψ angle of a randomly selected residue while a local move is to rotate seven successive torsion angles with keeping other residues unchanged. The step sizes of the global and local backbone movements are drawn from a Gaussian distribution with zero mean and standard deviation of 2° and 60°, respectively. A knowledge-based move formulated in previous research is to change the Φ and Ψ angles of the residue randomly to one of the 30 representative clustered points for each kind of amino acid (Chen, et al., 2007). Besides these backbone movements, a side chain move is to rotate all χ angles in a randomly selected nonproline residue. The step size of the side chain move is drawn from a Gaussian distribution with zero mean and standard deviation of 10°.

As shown in Figure 1B, in each step, the backbone move is carried out chosen from fragmove, knowledge-based move, local move and global move in order with the weights of 0.40, 0.33, 0.50 and 1.0, respectively. Energy is calculated and judged by the Metropolis criterion to decide whether this backbone move could be accepted followed by a round of side chain move and energy calculation.

Initialized structures were generated by the unfolding simulation without fragmove starting from the native structures at a very high temperature (t=1000). Fifteen random coil structures were constructed for each target protein. To optimize the weight of fragmove as well as the standard deviations of the step sizes of Φ and Ψ angles, we run REMC simulations to sample conformational space with 30 replicas at different temperatures, ranging from 0.15 to 1.50. Nine simulations were run for each parameter and each simulation has 50,000,000 steps. The optimized weight of fragmove is 0.40 while the optimized standard deviations of the step sizes of Φ and Ψ angles are 7.5° and 5.0°, respectively. To evaluate the performance of fragmove, for each protein in the training, testing and CASP12 sets, starting from different random coil structures, fifteen simulations were run with or without fragmove, respectively. Each simulation has 100,000,000 steps with 30 replicas at different temperatures ranging from 0.15 to 1.50. The trajectories at the lowest temperature (t=0.150) were analyzed for structure prediction. Secondary structures of the predicted structures and native structures were assigned by DSSP (Kabsch and Sander, 1983). The accuracy of predicted tertiary structure models were evaluated by their RMSD and TM-Score values.

## 3 Results and discussion

3.1 Improvement of secondary and tertiary structure

We built fragment libraries for each protein in the training set, testing set and CASP12 set by LRFragLib. Figure S1 shows the satisfactory performance of all the three datasets when evaluating precision and coverage of fragment libraries. More specifically, at the cutoff of 1.5Å, the training, testing and CASP12 set all show high levels of precision (67.66%, 83.86% and 65.40% respectively), while all of these three datasets achieve a coverage of >90%. These results indicates LRFragLib is powerful to identify near-native fragments and the fragment libraries constructed by LRFragLib have satisfactory qualities that can be well used for designing fragmove and thus contribute to ab initio protein structure prediction.

**Fig. 2.** Simulation results for the training, testing and CASP12 sets.

We run 15 simulations for each protein with or without fragmove, respectively and chose the minimum energy structures as the final models. When using fragmove, a lower RMSD value was observed on all proteins in the training set, 10/12 proteins in the testing set and 10/12 proteins in the CASP12 set (Figure 2A and Table S4&S5). Specifically, REMC simulations with fragmove successfully folded 14 proteins to moderate resolution

(RMSD<5 Å) and 6 proteins to high resolution (RMSD<3 Å) with native-like fold and topology while simulations without fragmove only fold 8 proteins to moderate resolution and 2 proteins to high resolution in all three datasets (Table S4&S5). On average, when using fragmove, the lowest RMSD value of the lowest energy structures were decreased by 2.55 Å, 1.35 Å and 1.95 Å in the testing set, CASP12 set and all tested proteins (including all proteins in the testing and CASP12 set), respectively.

Since RMSD is sensitive to outliers, we also made evaluation based on the TM-Scores of lowest energy structures. Similar results to evaluation based on RMSD, when using fragmove in REMC simulations, the highest and average TM-Score values of the energy minimum structures, for nearly all proteins in the training, testing and CASP12 sets, were higher than the numbers without fragmove (Figure 2 B&C and Table S6&S7). More specifically, the highest TM-Score observed increased by 23.97%, 7.06% and 15.52% while the average of TM-Score of the lowest energy structures decreased by 26.24%, 9.73% and 17.98% in the testing set, CASP12 set and all tested proteins, respectively. This results indicates that using fragmove in REMC simulations does not only improve the quality of the best model but also increase the overall accuracy of all simulation results. In addition, compared with CASP12 set, testing set achieved higher precision and coverage due to much more low homology proteins detected and thus it lead to the improvement on the testing set, although the improvements on both sets are statistical significant. In addition, using fragmove does not only improve the prediction accuracy of tertiary structures, but also contributes to the accuracy of secondary structures. Figure 2D and Table S8 show that 9/12 proteins in the tseting set and 11/12 proteins in CASP12 set achieved higher secondary structure accuracy when using fragmove in REMC simulation. Specifically, the accuracy was increased by 12.91%, 9.56% and 11.24% in the testing set, CASP12 set and all tested proteins, respectively.

**Fig. 3.** Superposition of energy minimum structures from 15 REMC simulations without fragmove (left

panel) or with fragmove (right panel).

Besides Figure 2 paints a comprehensive view of the improvement due to fragmove set, we also examined the lowest energy structures with the highest TM-Scores for each protein to gauge how fragmove makes such significant improvement. In Figure 3A on the left hand side, the energy minimum structure of the protein 1SHF from simulations without fragmove is imposed over the experimental PDB structure(green). 1SHF is a β protein with the length of 59 residues, however, the predicted structure without fragmove has a helix and thus the topology is incorrect. In comparison, in Figure 3A on the right hand side, the energy minimum structure (blue) from simulation with fragmove successfully predicted all four β strands in 1SHF and these strands also packed better to yield a much lower RMSD structure. Figure 3B shows the protein with a complex structure in CASP12 set, T0868-D1, which has five helices and five strands with the length of 116 residues. The structure without fragmove failed to predict the overall topology while the one with fragmove has the correct topology of four helices and two strands. Moreover, in Figure S2, predicted structure of 1AIL with fragmove shows excellent agreement with the native structure while structure without fragmove failed to predict the third helix. Although structures without/with fragmove both successfully predicted the topology of 1IGD and 1K9R in Figure S2, structures with fragmove packed much better and thus yield high TM-Scores and lower RMSD values. As for T0898-D1 in CASP12, the topology of the structure without fragmove is incorrect while the structure with fragmove correctly generated four helices and the topology was correct. These results indicates that using fragmove, on one hand, can yield more correctly folded secondary structures of alpha helices and beta strands, on the other hand, can improve the relative arrangement and packing between these secondary structures.

**Fig. 4.** Performance of the lowest energy.

## 3.2 Detection of lower energy conformation

We also discovered that in addition to being able to find structures much closer to the native structures, the fragmove set is also able to detect lower energy conformations. Figure 4 and Table S9-S10 show the lowest and average of energy values of the lowest energy structures for each target in the training, testing and CASP12 set. For all proteins in three datasets, predicted structures by using fragmove yielded lower energy with respect to the lowest and average values. The lowest energy was decreased by 4.62%, 3.84% and 4.23% while the average energy of lowest-energy structures was decreased by 5.12%, 5.16% and 5.72% in the testing set, CASP12 set and all tested proteins, respectively.

**Fig. 5.** The energy landscape for the proteins in ab initio REMC simulations with fragmove (gray dots) or without fragmove (black dots).

## 3.3 The energy landscape and improvement of efficiency

Since fragmove is able to detect lower RMSD and lower energy conformations, we next examined the nature of the energy landscape presented as energy-rmsd scatter plot (Figure 5). It can be seen that the free energy landscapes of simulations with fragmove were located on the lower left compared with that of simulations without fragmove, which indicates that conformations with fragmove found conformations with lower energy and structurally closer to native structures. Although the lowest energy conformations usually have quite low RMSD, the energies of native structures are not the lowest in all cases, even higher than many conformations during the simulations. This observation suggests that the energy function might not be accurate enough which mainly hinders to detect near-native structures. Interestingly, in most case, the conformation cluster with lowest RMSD values usually does not correspond to the one with lowest energies. Therefore, instead of simply selecting energy minimum structure as the final model, selection criteria of final protein structure as well as additional refinement steps to preserve and select the conformation clusters with lower RMSD should be discussed in the future work.

Besides detecting the energy landscape, we also analyzed the changes of energy and RMSD values during the folding process. Figure S3 shows the energy and RMSD time courses of 15 simulations of four protein cases in the testing and CASP12 sets. In the right hand panels, RMSD shows large discrete jumps due to the fact that in REMC, periodic exchanges of replica states between the lowest temperature (t=0.150) and higher temperatures. In each case, the energy and RMSD of simulations with fragmove both reached lower values faster than those of simulations without fragmove, especially for the free energy, which demonstrates that fragmove is also more efficient at guiding proteins to the low energy regions of conformational space.

## 4 Conclusion

We introduce an improved fragment-based movement, fragmove, which substitutes multiple backbone torsion angles with the values of known protein structures derived from fragment libraries generated by LRFragLib. Fragmove has been incorporated into REMC simulation and compared with a set of existing backbone movements in REMC, using fragmove yielded significant improved performance at increasing secondary structure and three dimensional model accuracy by 11.24% and 17.98%, respectively and reaching the energy minima decreased by 5.72%. Fragmove is more powerful to detect lower energy and structurally closer conformations to native structures, guide proteins much faster to low energy regions of conformational space and thus improve folding efficiency and protein structure prediction accuracy.

# Figures & Tables

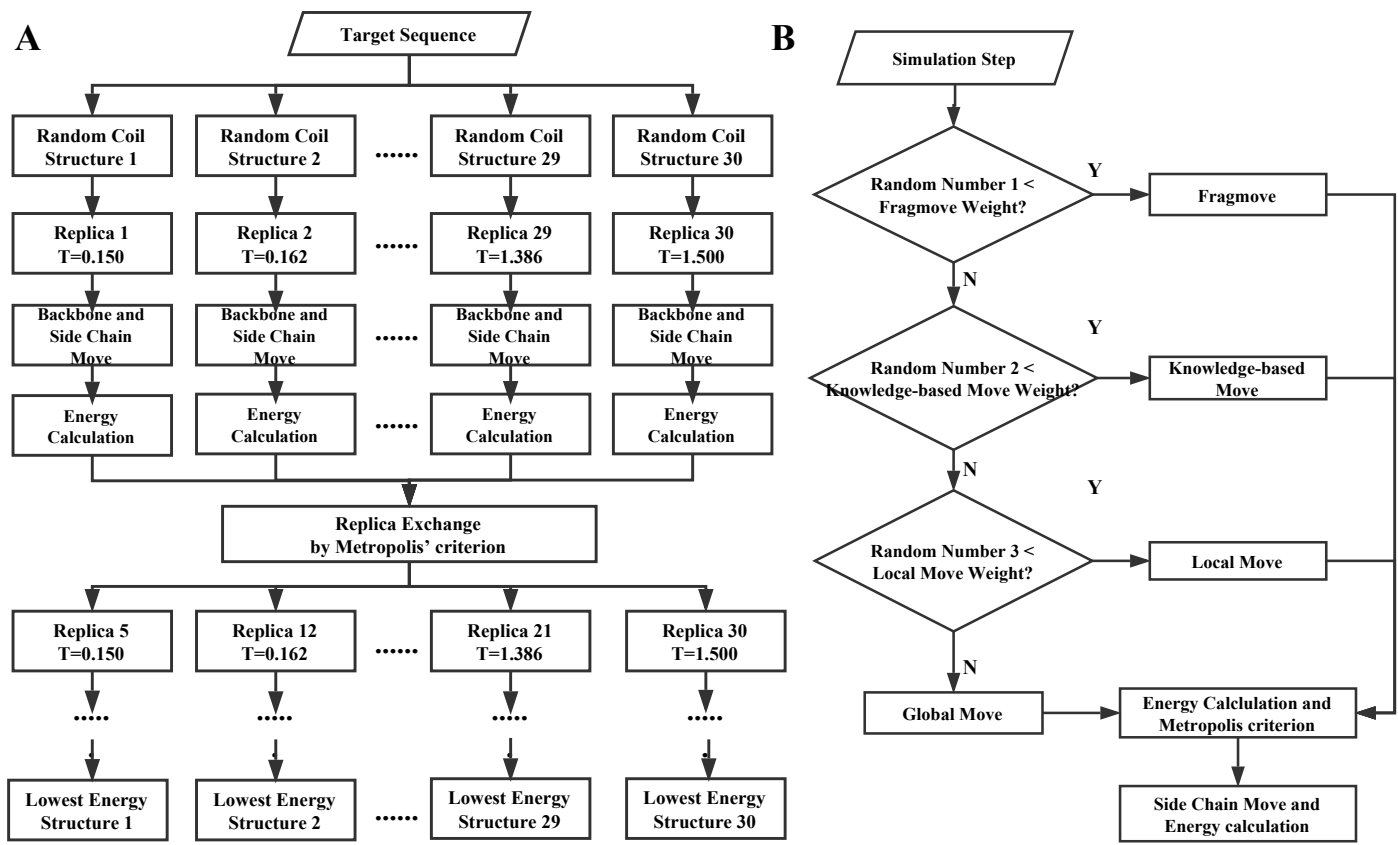

**Fig. 1.** General flowchart (A) and move sets in each step (B) of replica exchange Monte Carlo (REMC) simulation.

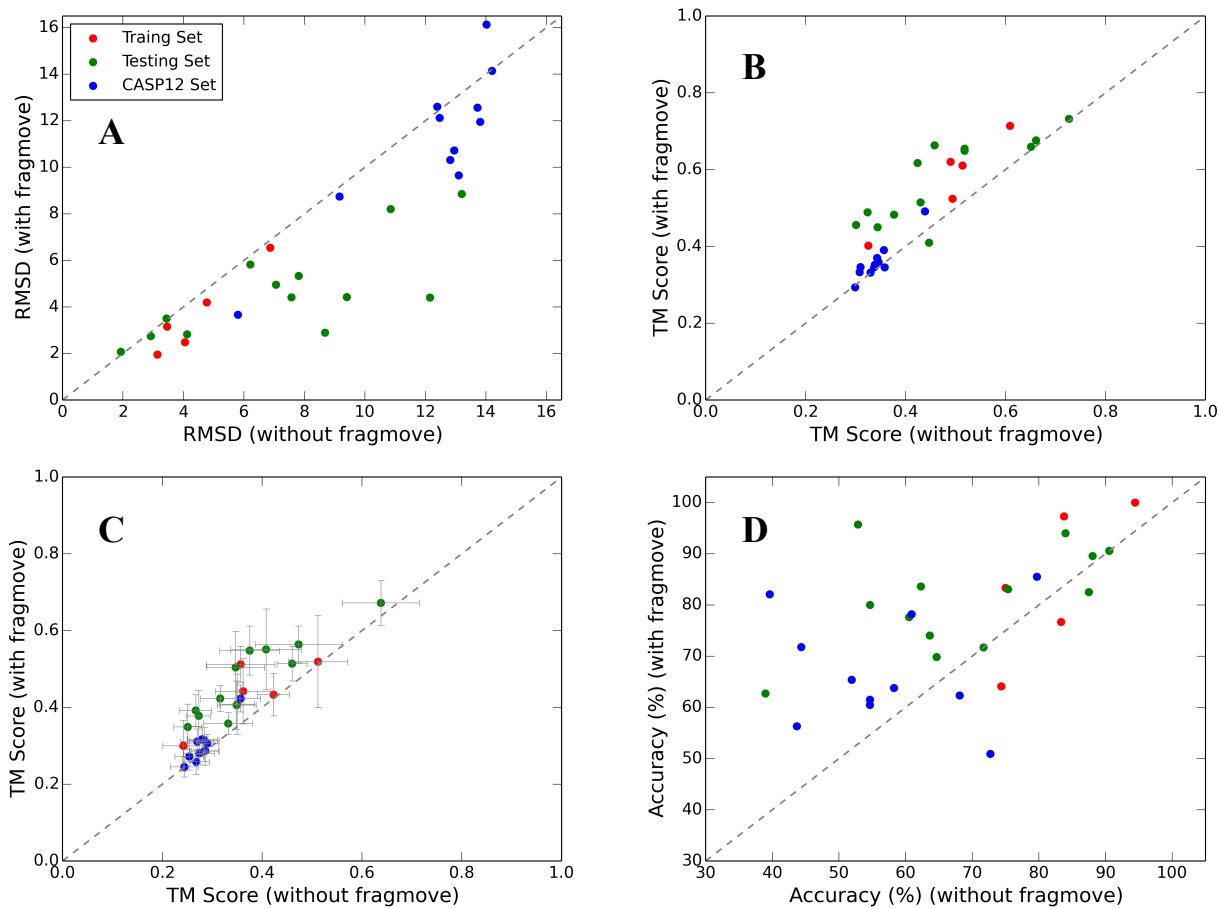

**Fig. 2.** Simulation results for the training, testing and CASP12 sets. Results were the lowest-energy structures obtained from fifteen trajectories at the lowest temperature (t=0.150) with/without fragmove. (A) Lowest RMSD values seen in fifteen lowest-energy structures. (B) Highest TM-Score values seen in fifteen lowest-energy structures. (C) Average and standard error of TM-Score values seen in fifteen lowest-energy structures. (D) Secondary structure accuracy of the highest TM-Score structure seen in fifteen lowest-energy structures.

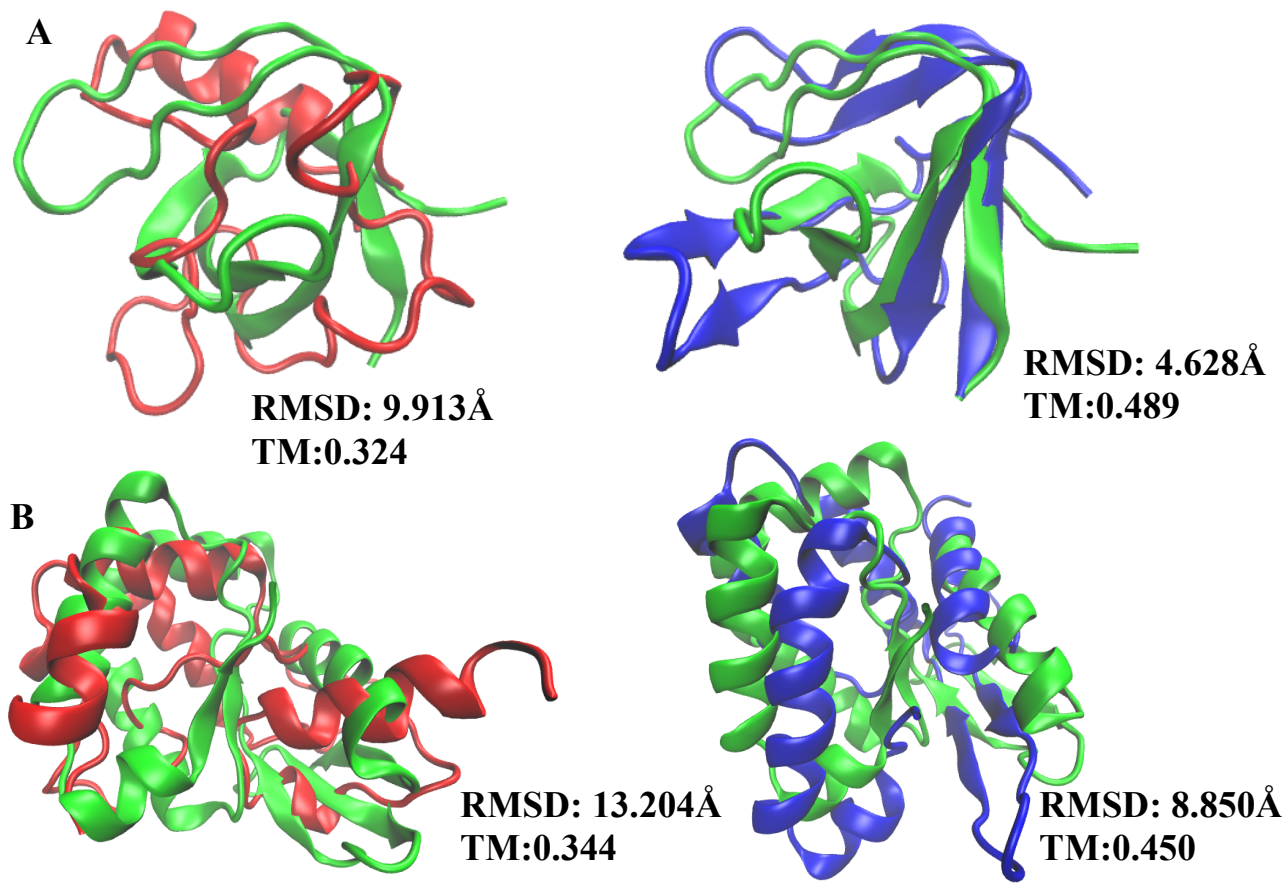

**Fig. 3.** Superposition of energy minimum structures from 15 REMC simulations without fragmove (left panel) or with fragmove (right panel); green: native structure, red: without fragmove, blue: with fragmove. (A) 1SHF, (B) T0868-D1.

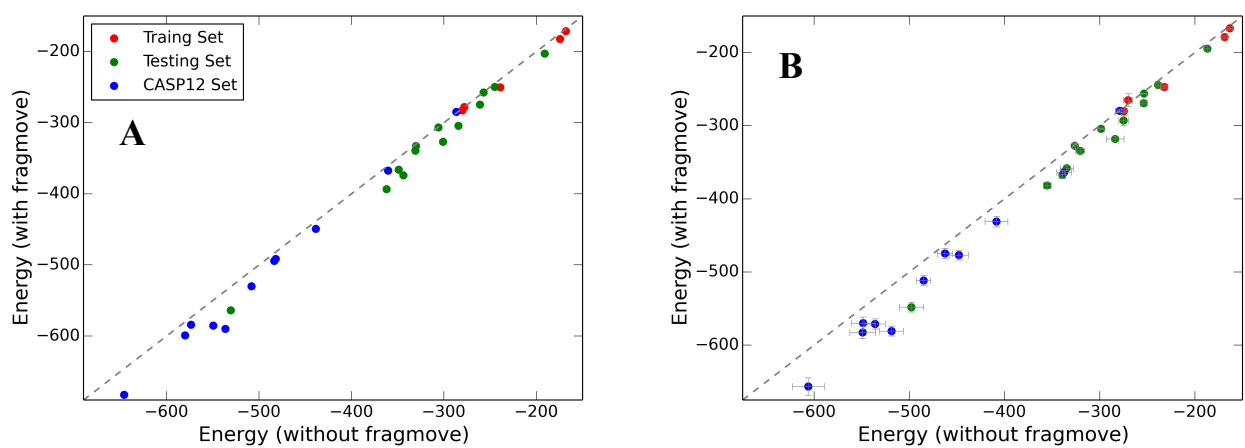

**Fig. 4.** Performance of the lowest energy (A) and the average and standard error of energy values (B) for the training, testing and CASP12 sets. Results were the lowest-energy structures obtained from fifteen trajectories at the lowest temperature (t=0.150) with/without fragmove.

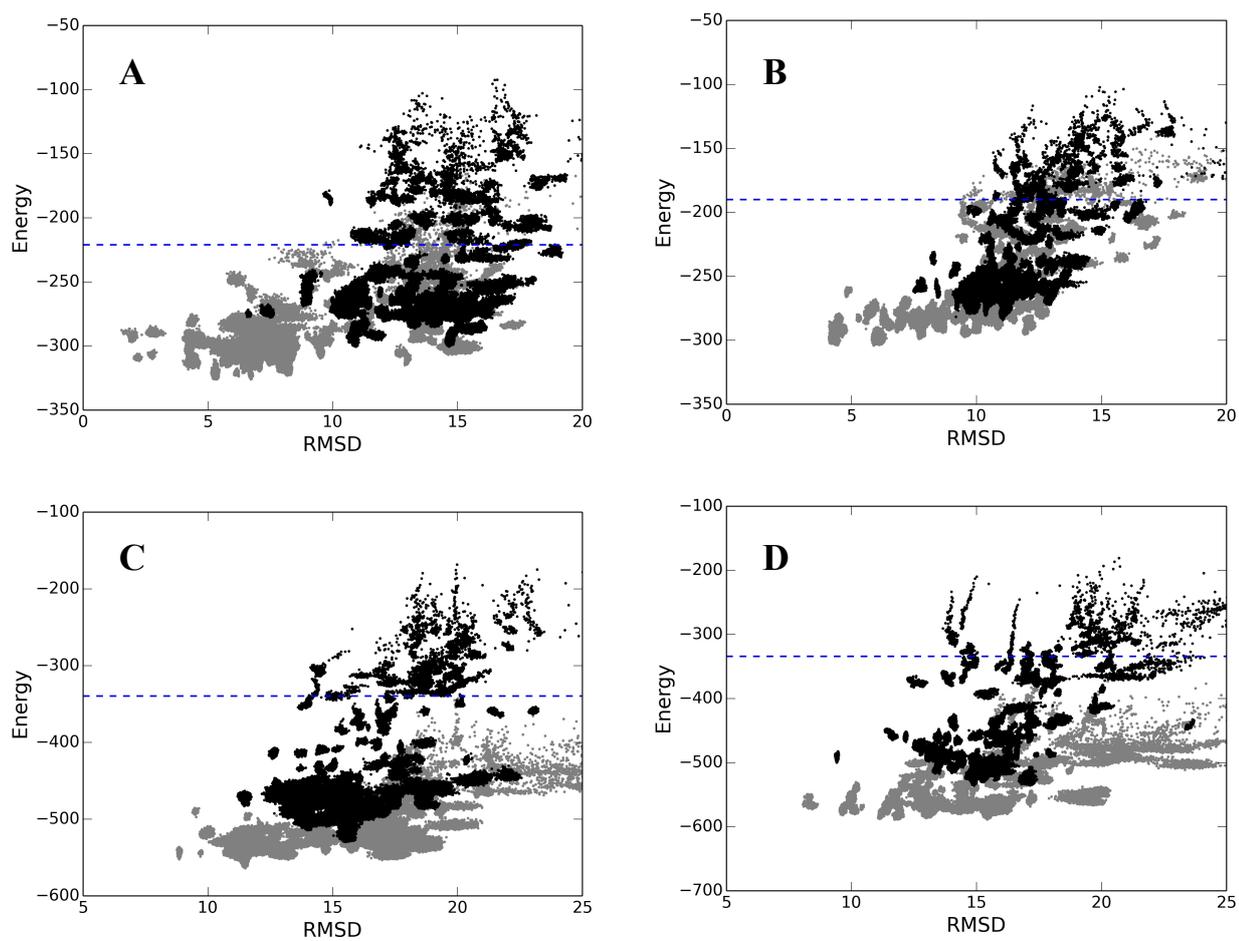

**Fig. 5.** The energy landscape for the proteins in ab initio REMC simulations with fragmove (gray dots) or without fragmove (black dots). (A) 1IGD, (B) 1SHF, (C) T0868-D1, (D) T0898-D1. Blue dotted line shows the energy of native structure.

# Supplementary Material

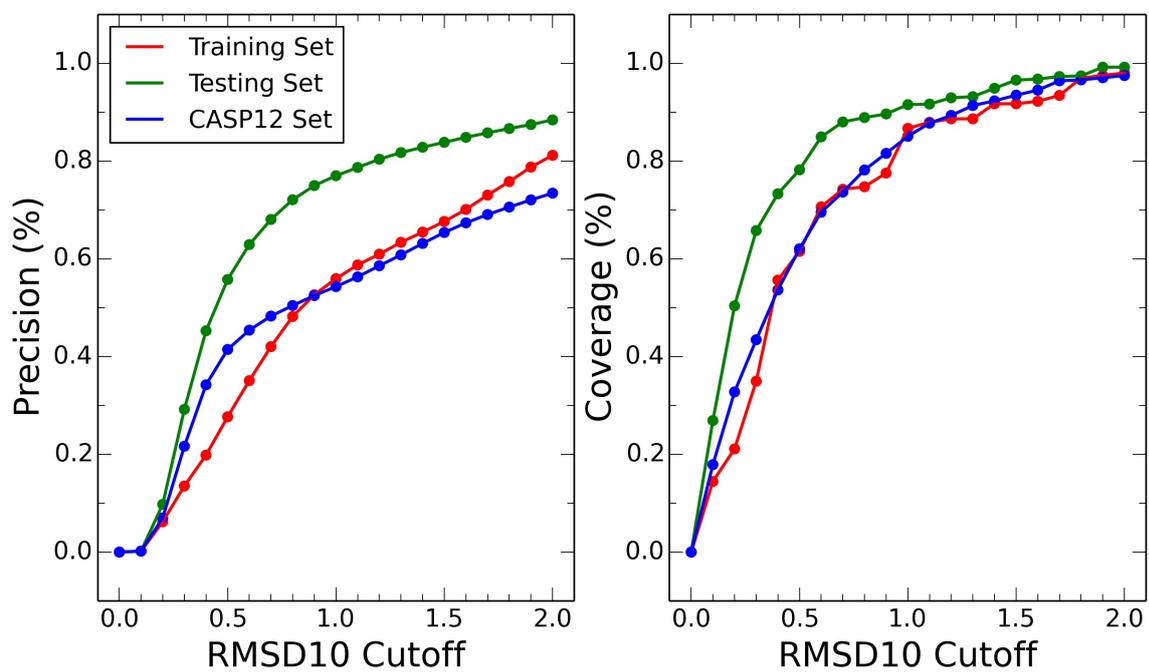

**Fig. S1. Performance of fragment libraries in the training, testing and CASP12 protein set.**

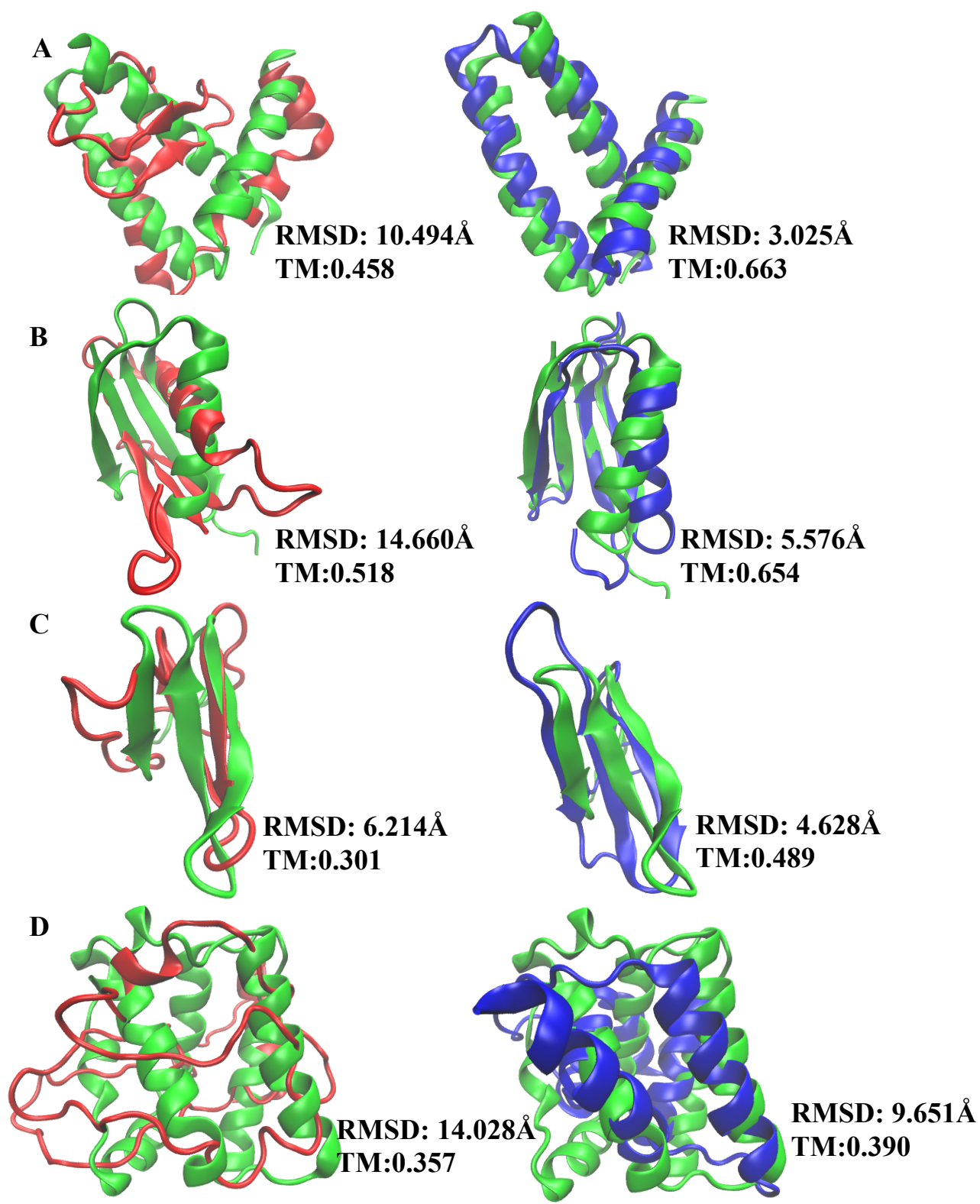

**Fig. S2.** Superposition of energy minimum structures from 15 REMC simulations without fragmove (left panel) or with fragmove (right panel); green: native structure, red: without fragmove, blue: with fragmove. (A) 1AIL, (B) 1IGD, (C) 1K9R, (D) T0898-D1.

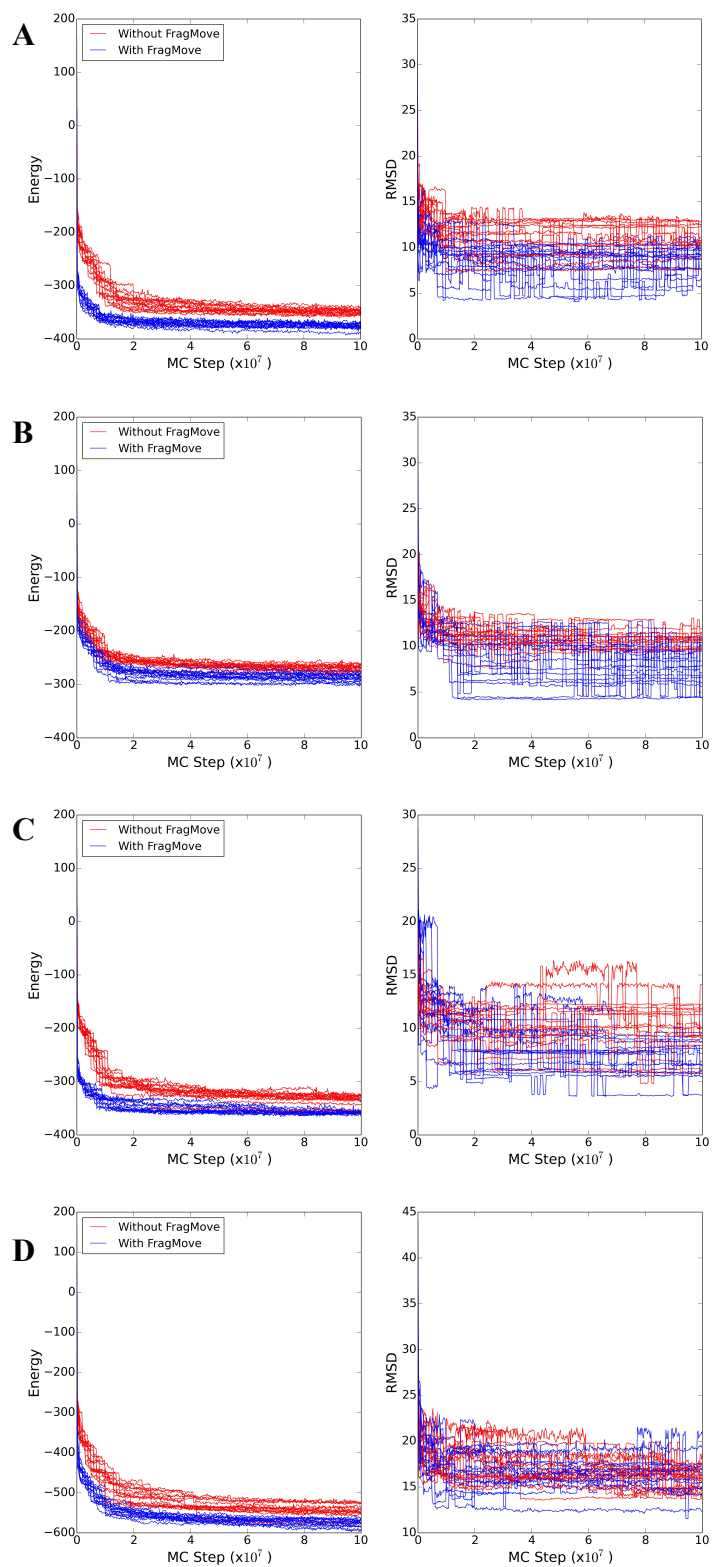

**Fig. S3.** Folding trajectories at the lowest temperature (t=0.150) of fifteen REMC simulations without fragmove (red lines) or with fragmove (blue lines). (A) 1CLB, (B) 1SHF, (C) T0892-D1, (D) T0897-D2.

**Table S1. Summary of proteins in the training set**

| Protein | Length | Fold |
|---------|--------|------|
| 1bdd    | 60     | α    |
| 1e0g    | 48     | αβ   |
| 1e0l    | 37     | β    |
| 1enh    | 54     | α    |
| 1i6c    | 39     | β    |

**Table S2. Summary of proteins in the testing set**

| Protein | Length | Fold |
|---------|--------|------|
| 1ail | 70 | α |
| 1ba5 | 53 | α |
| 1clb | 75 | αβ |
| 1gab | 53 | α |
| 1gjs | 65 | α |
| 1guu | 50 | α |
| 1igd | 61 | αβ |
| 1k9r | 40 | β |
| 1lfb | 77 | α |
| 1lq7 | 67 | α |
| 1shf | 59 | β |
| 1tif | 76 | αβ |

**Table S3. Summary of proteins in the CASP12 set**

| Protein   | Length | Fold |
|-----------|--------|------|
| T0868-D1  | 116    | αβ   |
| T0869-D1  | 104    | αβ   |
| T0886-D2  | 127    | αβ   |
| T0892-D1  | 69     | α    |
| T0892-D2  | 110    | αβ   |
| T0896-D1  | 86     | β    |
| T0896-D3  | 161    | αβ   |
| T0897-D1  | 138    | αβ   |
| T0897-D2  | 124    | αβ   |
| T0898-D1  | 106    | α    |
| T0898-D2  | 55     | β    |
| T0912-D3  | 103    | αβ   |

## Table S4. Summary of RMSD evaluation of Simulation Results for the Training and Testing Set

| Protein | $R_{min}$ | $R_{Emin}$ | $R_{min}^{frag}$ | $R_{Emin}^{frag}$ | $\Delta(R_{Emin}^{frag}-R_{Emin})$ |
|---------|-----------|------------|------------------|-------------------|-------------------------------------|
| 1bdd    | 2.90      | 4.77       | 3.34             | 4.19              | -0.58                               |
| 1e0g    | 3.14      | 4.05       | 2.38             | 2.48              | -1.57                               |
| 1e0l    | 2.98      | 3.46       | 2.42             | 3.15              | -0.31                               |
| 1enh    | 2.28      | 3.14       | 1.68             | 1.95              | -1.19                               |
| 1i6c    | 5.16      | 6.87       | 4.77             | 6.54              | -0.33                               |
| 1ail    | 6.85      | 8.68       | 2.63             | 2.89              | -5.79                               |
| 1ba5    | 5.34      | 7.06       | 4.66             | 4.95              | -2.11                               |
| 1clb    | 7.16      | 7.81       | 4.03             | 5.33              | -2.48                               |
| 1gab    | 2.52      | 2.92       | 2.35             | 2.74              | -0.18                               |
| 1gjs    | 3.01      | 3.44       | 2.94             | 3.50              | 0.06                                |
| 1guu    | 3.78      | 4.12       | 2.25             | 2.82              | -1.30                               |
| 1igd    | 6.51      | 7.57       | 1.53             | 4.41              | -3.16                               |
| 1k9r    | 5.92      | 6.21       | 5.05             | 5.82              | -0.39                               |
| 1lfb    | 8.00      | 12.15      | 3.90             | 4.40              | -7.75                               |
| 1lq7    | 1.74      | 1.93       | 1.89             | 2.07              | 0.14                                |
| 1shf    | 7.48      | 9.40       | 4.10             | 4.42              | -4.98                               |
| 1tif    | 8.59      | 10.85      | 5.12             | 8.20              | -2.65                               |

$R_{min}$：lowest RMSD values seen in fifteen simulations without fragmove.

$R_{Emin}$: lowest RMSD values seen in fifteen lowest-energy structures without fragmove.

$R_{min}^{frag}$ ：lowest RMSD values seen in fifteen simulations with fragmove.

$R_{Emin}^{frag}$ : lowest RMSD values seen in fifteen lowest-energy structures with fragmove.

All the lowest-energy structures were obtained from fifteen trajectories at the lowest temperature (t=0.150).

**Table S5. Summary of RMSD evaluation of Simulation Results for the CASP12 Testing Set**

| Protein | $R_{min}$ | $R_{Emin}$ | $R_{min}^{frag}$ | $R_{Emin}^{frag}$ | $\Delta(R_{Emin}^{frag}-R_{Emin})$ |
|---------|-----------|------------|------------------|-------------------|-------------------------------------|
| T0868-D1 | 11.12 | 13.20 | 8.75 | 8.85 | -4.35 |
| T0869-D1 | 11.61 | 12.47 | 11.34 | 12.12 | -0.35 |
| T0886-D2 | 12.96 | 14.20 | 11.50 | 14.14 | -0.06 |
| T0892-D1 | 4.74 | 5.80 | 3.52 | 3.66 | -2.14 |
| T0892-D2 | 12.42 | 12.82 | 10.06 | 10.31 | -2.51 |
| T0896-D1 | 10.49 | 12.39 | 10.09 | 12.6 | 0.21 |
| T0896-D3 | 13.66 | 14.02 | 13.98 | 16.13 | 2.11 |
| T0897-D1 | 11.81 | 13.81 | 11.07 | 11.95 | -1.86 |
| T0897-D2 | 13.50 | 13.72 | 11.48 | 12.56 | -1.16 |
| T0898-D1 | 9.34 | 13.10 | 8.01 | 9.65 | -3.45 |
| T0898-D2 | 7.11 | 9.16 | 8.47 | 8.74 | -0.42 |
| T0912-D3 | 11.90 | 12.95 | 10.38 | 10.72 | -2.23 |

$R_{min}$: lowest RMSD values seen in fifteen simulations without fragmove.

$R_{Emin}$: lowest RMSD values seen in fifteen lowest-energy structures without fragmove.

$R_{min}^{frag}$ : lowest RMSD values seen in fifteen simulations with fragmove.

$R_{Emin}^{frag}$ : lowest RMSD values seen in fifteen lowest-energy structures with fragmove.

All the lowest-energy structures were obtained from fifteen trajectories at the lowest temperature (t=0.150).

**Table S6. Summary of TM-Score evaluation of Simulation Results for the Training and Testing Sets**

| Protein | $TM_{Emin}$ | $TM_{Emin}^{frag}$ | $\Delta(TM_{Emin}^{frag}-TM_{Emin})$ | Percentage (%) | $TM_{Ave}$ | $TM_{Ave}^{frag}$ | $\Delta(TM_{Ave}^{frag}-TM_{Ave})$ | Percentage (%) |
|---|---|---|---|---|---|---|---|---|
| 1bdd | 0.494 | 0.524 | 0.030 | 6.05 | 0.423 | 0.433 | 0.010 | 2.36 |
| 1e0g | 0.514 | 0.611 | 0.097 | 18.78 | 0.357 | 0.512 | 0.155 | 43.42 |
| 1e0l | 0.490 | 0.620 | 0.130 | 26.58 | 0.362 | 0.442 | 0.080 | 22.10 |
| 1enh | 0.609 | 0.714 | 0.105 | 17.21 | 0.512 | 0.519 | 0.007 | 1.37 |
| 1i6c | 0.326 | 0.402 | 0.076 | 23.43 | 0.242 | 0.300 | 0.058 | 23.97 |
| 1ail | 0.458 | 0.663 | 0.205 | 44.77 | 0.347 | 0.504 | 0.157 | 45.24 |
| 1ba5 | 0.377 | 0.483 | 0.106 | 28.00 | 0.316 | 0.423 | 0.107 | 33.86 |
| 1clb | 0.430 | 0.514 | 0.084 | 19.64 | 0.349 | 0.406 | 0.057 | 16.33 |
| 1gab | 0.661 | 0.676 | 0.015 | 2.30 | 0.408 | 0.551 | 0.143 | 35.05 |
| 1gjs | 0.651 | 0.659 | 0.008 | 1.28 | 0.473 | 0.564 | 0.091 | 19.24 |
| 1guu | 0.518 | 0.649 | 0.131 | 25.28 | 0.460 | 0.514 | 0.054 | 11.74 |
| 1igd | 0.518 | 0.654 | 0.136 | 26.31 | 0.375 | 0.548 | 0.173 | 46.13 |
| 1k9r | 0.301 | 0.456 | 0.155 | 51.38 | 0.251 | 0.349 | 0.098 | 39.04 |
| 1lfb | 0.424 | 0.617 | 0.193 | 45.55 | 0.350 | 0.408 | 0.058 | 16.57 |
| 1lq7 | 0.727 | 0.732 | 0.005 | 0.68 | 0.638 | 0.672 | 0.034 | 5.33 |
| 1shf | 0.324 | 0.489 | 0.165 | 50.86 | 0.273 | 0.378 | 0.105 | 38.46 |
| 1tif | 0.447 | 0.410 | -0.037 | -8.38 | 0.332 | 0.358 | 0.026 | 7.83 |

$TM_{Emin}$ : highest TM-Score values seen in fifteen lowest-energy structures without fragmove.
$TM_{Emin}^{frag}$ : highest TM-Score values seen in fifteen lowest-energy structures with fragmove.
$TM_{ave}$ : the average of TM-Score values of fifteen lowest-energy structures without fragmove.
$TM_{ave}^{frag}$ : the average of TM-Score values of fifteen lowest-energy structures with fragmove.
All the lowest-energy structures were obtained from fifteen trajectories at the lowest temperature (t=0.150).

### Table S7. Summary of TM-Score evaluation of Simulation Results for the CASP12 Testing Sets

| Protein | $TM_{Emin}$ | $TM_{Emin}^{frag}$ | $\Delta(TM_{Emin}^{frag}-TM_{Emin})$ | Percentage(%) | $TM_{Ave}$ | $TM_{Ave}^{frag}$ | $\Delta(TM_{Ave}^{frag}-TM_{Ave})$ | Percentage(%) |
|---|---|---|---|---|---|---|---|---|
| T0868-D1 | 0.344 | 0.450 | 0.106 | 30.76 | 0.267 | 0.392 | 0.125 | 46.82 |
| T0869-D1 | 0.337 | 0.346 | 0.008 | 2.47 | 0.290 | 0.306 | 0.016 | 5.52 |
| T0886-D2 | 0.346 | 0.359 | 0.013 | 3.70 | 0.270 | 0.311 | 0.041 | 15.19 |
| T0892-D1 | 0.439 | 0.491 | 0.052 | 11.93 | 0.357 | 0.423 | 0.066 | 18.49 |
| T0892-D2 | 0.343 | 0.370 | 0.027 | 7.83 | 0.284 | 0.314 | 0.030 | 10.56 |
| T0896-D1 | 0.308 | 0.333 | 0.025 | 8.00 | 0.254 | 0.272 | 0.018 | 7.09 |
| T0896-D3 | 0.299 | 0.293 | -0.006 | -1.95 | 0.244 | 0.245 | 0.001 | 0.41 |
| T0897-D1 | 0.338 | 0.352 | 0.013 | 3.95 | 0.284 | 0.286 | 0.002 | 0.70 |
| T0897-D2 | 0.310 | 0.346 | 0.036 | 11.68 | 0.268 | 0.258 | -0.010 | -3.73 |
| T0898-D1 | 0.357 | 0.390 | 0.034 | 9.42 | 0.280 | 0.317 | 0.037 | 13.21 |
| T0898-D2 | 0.358 | 0.345 | -0.013 | -3.60 | 0.286 | 0.287 | 0.001 | 0.35 |
| T0912-D3 | 0.330 | 0.332 | 0.002 | 0.53 | 0.274 | 0.28 | 0.006 | 2.19 |

$TM_{Emin}$ : highest TM-Score values seen in fifteen lowest-energy structures without fragmove.

$TM_{Emin}^{frag}$ : highest TM-Score values seen in fifteen lowest-energy structures with fragmove.

$TM_{ave}$ : the average of TM-Score values of fifteen lowest-energy structures without fragmove.

$TM_{ave}^{frag}$ : the average of TM-Score values of fifteen lowest-energy structures with fragmove.

All the lowest-energy structures were obtained from fifteen trajectories at the lowest temperature (t=0.150).

**Table S8. Summary of Secondary Structure Accuracy of Simulation Results for Training, Testing and CASP12 Testing Sets**

| Protein | Acc_TMEmin (%) | Acc_TMfragEmin (%) | Protein | Acc_TMEmin (%) | Acc_TMfragEmin (%) |
|---|---|---|---|---|---|
| 1bdd | 83.33 | 76.67 | 1shf | 38.98 | 62.71 |
| 1e0g | 75.00 | 83.33 | 1tif | 60.53 | 77.63 |
| 1e0l | 83.78 | 97.30 | T0868-D1 | 64.66 | 69.83 |
| 1enh | 94.44 | 100.00 | T0869-D1 | 51.92 | 65.38 |
| 1i6c | 74.36 | 64.10 | T0886-D2 | 58.27 | 63.78 |
| 1ail | 52.86 | 95.71 | T0892-D1 | 79.71 | 85.51 |
| 1ba5 | 71.70 | 71.70 | T0892-D2 | 60.91 | 78.18 |
| 1clb | 54.67 | 80.00 | T0896-D1 | 54.65 | 60.47 |
| 1gab | 90.57 | 90.57 | T0896-D3 | 54.66 | 61.49 |
| 1gjs | 75.38 | 83.08 | T0897-D1 | 68.12 | 62.32 |
| 1guu | 84.00 | 94.00 | T0897-D2 | 44.35 | 71.77 |
| 1igd | 62.30 | 83.61 | T0898-D1 | 39.62 | 82.08 |
| 1k9r | 87.50 | 82.50 | T0898-D2 | 72.73 | 50.91 |
| 1lfb | 63.64 | 74.03 | T0912-D3 | 43.69 | 56.31 |
| 1lq7 | 88.06 | 89.55 | | | |

Acc_TMEmin: secondary structure accuracy of the highest TM-Score structure seen in fifteen lowest-energy structures without fragmove.

Acc_TMfragEmin: secondary structure accuracy of the highest TM-Score structure seen in fifteen lowest-energy structures with fragmove.

**Table S9. Summary of Energy evaluation of Simulation Results for the Training and Testing Sets**

| Protein | $E_{Emin}$ | $E_{Emin}^{frag}$ | $\Delta(E_{Emin}^{frag}-E_{Emin})$ | Percentage (%) | $E_{Ave}$ | $E_{Ave}^{frag}$ | $\Delta(E_{Ave}^{frag}-E_{Ave})$ | Percentage (%) |
|---|---|---|---|---|---|---|---|---|
| 1bdd | -280.10 | -282.81 | -2.71 | 0.97 | -274.48 | -280.12 | -5.64 | 2.05 |
| 1e0g | -239.03 | -250.36 | -11.33 | 4.74 | -232.09 | -247.13 | -15.04 | 6.48 |
| 1e0l | -168.08 | -171.35 | -3.27 | 1.95 | -163.21 | -166.90 | -3.69 | 2.26 |
| 1enh | -277.91 | -278.10 | -0.19 | 0.07 | -270.10 | -265.17 | 4.93 | -1.83 |
| 1i6c | -174.39 | -182.38 | -7.99 | 4.58 | -168.77 | -178.84 | -10.07 | 5.97 |
| 1ail | -330.87 | -339.45 | -8.58 | 2.59 | -320.45 | -334.46 | -14.01 | 4.37 |
| 1ba5 | -261.04 | -274.65 | -13.61 | 5.21 | -253.70 | -269.33 | -15.63 | 6.16 |
| 1clb | -362.04 | -393.58 | -31.54 | 8.71 | -355.08 | -381.77 | -26.69 | 7.52 |
| 1gab | -257.11 | -257.64 | -0.53 | 0.21 | -253.45 | -256.10 | -2.65 | 1.05 |
| 1gjs | -305.93 | -306.76 | -0.83 | 0.27 | -298.49 | -304.27 | -5.78 | 1.94 |
| 1guu | -245.08 | -249.79 | -4.71 | 1.92 | -238.68 | -244.26 | -5.58 | 2.34 |
| 1igd | -300.98 | -326.95 | -25.97 | 8.63 | -283.75 | -318.25 | -34.50 | 12.16 |
| 1k9r | -190.95 | -202.93 | -11.98 | 6.27 | -186.87 | -194.69 | -7.82 | 4.18 |
| 1lfb | -348.93 | -366.20 | -17.27 | 4.95 | -334.58 | -358.11 | -23.53 | 7.03 |
| 1lq7 | -330.23 | -332.83 | -2.60 | 0.79 | -326.16 | -327.46 | -1.30 | 0.40 |
| 1shf | -284.31 | -304.49 | -20.18 | 7.10 | -274.88 | -292.91 | -18.03 | 6.56 |
| 1tif | -343.87 | -374.13 | -30.26 | 8.80 | -339.18 | -367.11 | -27.93 | 8.23 |

$E_{Emin}$ : lowest energy values seen in fifteen lowest-energy structures without fragmove.

$E_{Emin}^{frag}$ : lowest energy values seen in fifteen lowest-energy structures with fragmove.

$E_{ave}$ : the average of energy values of fifteen lowest-energy structures without fragmove.

$E_{ave}^{frag}$ : the average of energy values of fifteen lowest-energy structures with fragmove.

All the lowest-energy structures were obtained from fifteen trajectories at the lowest temperature (t=0.150).

## Table S10. Summary of Energy evaluation of Simulation Results for the CASP12 Testing Sets

| Protein | $E^{Emin}$ | $E_{Emin}^{frag}$ | $\Delta(E_{Emin}^{frag} - E_{Emin})$ | Percentage (%) | EEAve | EfragEAve | $\Delta(E_{Ave}^{frag} - E_{Ave})$ | Percentage (%) |
|---|---|---|---|---|---|---|---|---|
| T0868-D1 | -530.65 | -563.96 | -33.31 | 6.28 | -497.76 | -548.20 | -50.44 | 10.13 |
| T0869-D1 | -508.20 | -530.12 | -21.92 | 4.31 | -484.85 | -511.72 | -26.87 | 5.54 |
| T0886-D2 | -549.58 | -585.43 | -35.85 | 6.52 | -535.91 | -571.13 | -35.22 | 6.57 |
| T0892-D1 | -360.40 | -367.51 | -7.11 | 1.97 | -337.26 | -363.70 | -26.44 | 7.84 |
| T0892-D2 | -483.57 | -494.47 | -10.90 | 2.25 | -462.38 | -474.71 | -12.33 | 2.67 |
| T0896-D1 | -438.59 | -449.37 | -10.78 | 2.46 | -408.40 | -431.02 | -22.62 | 5.54 |
| T0896-D3 | -645.99 | -682.73 | -36.74 | 5.69 | -606.00 | -656.58 | -50.58 | 8.35 |
| T0897-D1 | -573.61 | -584.22 | -10.61 | 1.85 | -548.44 | -569.92 | -21.48 | 3.92 |
| T0897-D2 | -580.06 | -599.20 | -19.14 | 3.30 | -548.95 | -582.64 | -33.69 | 6.14 |
| T0898-D1 | -536.44 | -590.03 | -53.59 | 9.99 | -518.53 | -580.82 | -62.29 | 12.01 |
| T0898-D2 | -286.66 | -284.93 | 1.73 | -0.60 | -278.97 | -279.69 | -0.72 | 0.26 |
| T0912-D3 | -481.96 | -491.73 | -9.77 | 2.03 | -447.82 | -476.92 | -29.10 | 6.50 |

$E_{Emin}$ : lowest energy values seen in fifteen lowest-energy structures without fragmove.

$E_{Emin}^{frag}$ : lowest energy values seen in fifteen lowest-energy structures with fragmove.

$E_{ave}$ : the average of energy values of fifteen lowest-energy structures without fragmove.

$E_{ave}^{frag}$ : the average of energy values of fifteen lowest-energy structures with fragmove.

All the lowest-energy structures were obtained from fifteen trajectories at the lowest temperature (t=0.150).